# β-KMS Green functions generating functionals
# - the convexity based approach

ROMAN GIELERAK
University of Zielona Gora
Institute of Control&Computations Engineering
r.gielerak@issi.uz.zgora.pl

## Abstract

*The notion of the stochastically positive β-KMS (Euclidean time) Green functionals on (the abelian sectors of) the CCR-algebras in Weyl form has been introduced. The main observation is that the essential properties of such functionals are stable against taking convex superpositions of them. Starting from the free Bose matter describing functionals (the condensed state situation is included as well) a constructive approach to the constructions of a new, of such type functionals is presented. In particular, starting from the free Bose matter Green functionals a class of not quasi-free thermal functionals, together with the corresponding modular structures have been constructed. Some elementary properties of models constructed are being presented.*

## 1. Introduction

One of the most important resource in quantum information processing technology are quantum correlations contained in entangled quantum states [1,2,3,4,5]. However, the phenomena of entanglement is extremely fragile as an arbitrary small perturbations is able to destruct this property[2,3,4,5].

In order to be able to use quantum entanglement effectively in genuine quantum technologies one has to look for a quantum entangled systems in which the quantum entanglements are more robust against different kinds of perturbations, like that caused for example, by hardly to be eliminated quantum noise of different origin. The systems in which the so called quantum macroscopic effects are present seems to be the proper alternative.The quantum behaviour on the macroscopic scale seems to form situation in which quantum correlations could survive on a much longer time-scale and thus allowing to perform realistic ,physical manipulations due to the process of executing basic quantum demands on them.

The Bose-Einstein condensates [5,6] seems to be an excellent and promising examples of quantum behaving systems of macroscopic size ,and therefore they might be considered as a good candidates for building quantum hardware for a future quantum computing machines [2,3,4,5 ].Phase transitions to quantum condensed phases—such as Bose–Einstein condensation (BEC), superfluidity, and superconductivity—have for a long fascinated macroscopic scientists, as they bring pure quantum effects to macroscale. BEC has, for example, famously been demonstrated in dilute atom gas of rubidium atoms at temperatures below 200 nanokelvin. Much effort has been devoted to finding a solid-state system in which BEC can take place. Promising candidate systems for this are semiconductor microcavities, in which photons are confined and strongly coupled to electronic excitations, leading to the creation of exciton polaritons. These bosonic quasi-particles are $10^9$ times lighter than rubidium atoms, thus theoretically permitting BEC to occur at standard cryogenic temperatures [6,7,8]**.**

The present note includes some mathematical construction and tools for the construction of a new, large classes of mathematically well treated,selfinteracting Bose matter models starting from some quasi-free models. The standard, free , critical Bose gas being in the condensed state is among them , however the corresponding constructions and analysis of this case will presented in a separate paper [10].Detailed studies of the models constructed here, especially from the point of view of entanglement present in the constructed quantum states will be presented in an another paper of the author [10 ].

In the paper [20] very similar ideas were applied for a constructions of a class of of interacting Wightman quantum fields in any dimensions .

## 2.Thermal , β-KMS Euclidean- time Green functionals.

The notion of the stochastically positive β-KMS structures, like thermal states, Green functions,…etc. on a abelian $C^*$-algebras has been originated in [ 11,12 ] and developed then in papers [13-19] to which we are referring the interested reader for more details.

The starting point of the present note is the notion of a thermal stochastic process $\zeta_t$ taking values in some , separable Hilbert space H.Of special interest will be the case of infinite dimensional space H, in particular the case H=$L^2(R^d)$ , d$\geq$ 1 .The process $\zeta_t$ : $K_\beta \rightarrow$ H , where $K_\beta$ is the circle , modelled as the interval [-β/2,β/2 ] with endpoints glued is fully described by its path space measure on the space of trajectories { t ε $K_\beta \rightarrow \zeta_t$ ε H } .Under certain regularity assumption , an appropriate nuclear rigging procedure can be

*Bose Matter models*

used to localise the path space measure of the process $\zeta_t$ as the Borel, cylindric measure on the space $H^*$ in the nuclear rigging constructed $H_* \to H \to H^*$. In the case of $H=L^2(R^d)$ the standard Schwartz nuclear rigging $D(R^d) \to L^2(R^d) \to D^*(R^d)$ or $S(R^d) \to L^2(R^d) \to S^*(R^d)$ will be used.

The Euclidean time, Green functions G of the thermal process $\zeta_t$ are defined as:

$$G\left((\tau_1, f_1)\ldots,(\tau_n, f_n)\right) =$$
$$E\prod_{\alpha=1}^n (\exp(i<\zeta_{\tau_\alpha}, f_\alpha>) = \quad (2.1)$$
$$\int_{H^*} d\mu^\zeta(\Phi) \prod_{\alpha=1}^n \exp(i(<\Phi, \delta_{\tau_\alpha} \otimes f_\alpha>))$$

where $\tau_i \in K_\beta$, $f_i \in H_*$.

The β-KMS Green functional of the thermal process $\zeta_\tau \in H$ is given by:

$$\mathbf{G}^\beta(\psi \otimes f) = E\left(\exp\left(i\int_{-\beta/2}^{\beta/2} d\tau \psi(\tau) <\zeta_\tau, f>\right)\right) \quad (2.2)$$

for ψ being real valued, continuous function on the circle $K_\beta$ and $f \in H_*$. It is not difficult to see that the knowledge of $\mathbf{G}^\beta$ enables us to compute all the Green functions (2.1) of the process $\zeta_\tau$.

From the very definition of the thermal process the following result follows.

**Theorem 2.1**

*Let $\zeta_\tau : K_\beta \to H$ be a thermal process and let $(H_*, H^*)$ be the corresponding to the assumed regularity of the process, nuclear rigging of the Hilbert space H. Then the Green functional $\mathbf{G}^\beta$ of the process $\zeta_\tau$ obeys the following properties:*

**G(1)**

*the functional $\mathbf{G}^\beta$ is S-positive, i.e. for any sequences $c_\alpha \in C$, $\psi_\alpha \in C(K_\beta)$, $f_\alpha \in H_*$, $\alpha=1:n$,*

$$\sum_{\alpha,\gamma} c_\alpha^- * c_\gamma * \mathbf{G}(\psi_\alpha \otimes f_\alpha - \psi_\gamma \otimes f_\gamma) \geq 0 \quad (2.3)$$

**G(2)**

*the functional $\mathbf{G}^\beta$ is reflection, R-positive, i.e. i.e. for any sequences $c_\alpha \in C$, $\psi_\alpha \in C([0,\beta/2])$, $f_\alpha \in H_*$, $\alpha=1:n$,*

$$\sum_{\alpha,\gamma} c_\alpha^- * c_\gamma * G(\psi_\alpha \otimes f_\alpha - R\psi_\gamma \otimes f_\gamma) \geq 0 \quad (2.4)$$
*(2.4)*

*where R is the reflection operator on $K_\beta$.*

**G(3)**

*the functional $\mathbf{G}^\beta$ is reflection invariant, $K_\beta$ shift-invariant and β-periodic.*

Under certain regularity assumption on ( see [10]  for details ) **G**$^\beta$ together with **G(1) – G(3)** the following inverse result is valid :

**Theorem 2.2 .**
*Let ( H$_*$ , H$^*$ ) be  a nuclear rigging of the separable Hilbert  space H and let **G**$^\beta$ be a nonlinear functional on $C_r( K_\beta )\otimes H^*$ obeying  G(1) –G(3) and  some additional regularity condition ( see [ 10 ] ) .Then , there exists an unique ( up to the stochastic equivalence ) thermal process $\zeta_\tau \epsilon H^*$, such that the Green functional of which is equal to **G**$^\beta$ .*

In the particular situation of $H = L_2( R^d )$ , for $d \geq 1$ and for the standard nuclear rigging $S( R^d ) \to L_2( R^d ) \to S'( R^d )$ we can construct , on the corresponding Abelian subalgebra $A(L_2( R^d ))$ of the Weyl algebra $W(L_2( R^d ))$ from a given **G**$^\beta$ the corresponding system of multitime ,Euclidean Green functions obeying the system of axioms proposed in [18] .Therefore , applying the construction presented in [ 18] we can prove the following theorem.

**Theorem 2.3.**
*Let $S( R^d )\to L_2( R^d ) \to S'( R^d )$ be a standard nuclear rigging of $L_2( R^d )$ and let **G**$^\beta$ be  a nonlinear  functional  on  $C_r( K_\beta )\otimes H_*$  obeying  G(1) –G(3) and  appropriate temperedness condition. Then , there exists a β-KMS  structure (R, $\alpha_t , \omega^\beta$ ), such that the corresponding , Euclidean - time Green functions of which*

$$G((\tau_1, f_1), \ldots, (\tau_n, f_n)) = \omega^\beta \left( \prod_{k=1}^n \alpha_{i\tau_k} \left( \pi(W(f_k)) \right) \right) \qquad (2.5)$$

*where π is the corresponding *- representation of the Weyl algebra ( an abelian part of W((h) )and R is *-closure of the image of $\alpha_{i\tau}$ $(A(R^d))$ under π.*

The basic observation for our constructions illustrated in the next section
is that the fundamental properties **G(1)-G(3)** of thermal funcionals are stable under taking
suitable convex superposition operations.

**Proposition 2.4.**

*Let (Ω, Σ,dP ) be some probablistic space and let **G**$^\beta$ (ω, .) be a weakly Σ measurable family of thermal β-KMS Green functionals obeying properties G(1) – G(3) together with the uniform regularity condition a. e. on Ω .Then the functional*

$$\mathbf{G}^\beta{}_P (\psi \otimes f) = \int_\Omega dP(\omega)\, \mathbf{G}^\beta(\omega, \psi \otimes f))$$

(2.6)

*Bose Matter models*

*obeys G(1)- G(3).*

Let us define also, the corresponding to the thermal process $\varsigma_\tau$ path space measure $d\mu_\varsigma(\varphi)$ Schwinger functions

$$S^\beta((\tau_1,f_1),\ldots,(\tau_n,f_n)) = \int_{H^*} d\mu_\varsigma(\varphi) \prod_{\alpha=1}^n <\varphi, \delta_{\tau_\alpha} \otimes f_\alpha>$$

$$= \frac{1}{j^n} \frac{\delta^n}{\delta_{t_1}\ldots\delta_{t_n}} (G^\beta((\tau_1, t_1 f_1)\ldots,(\tau_n, t_n f_n)))\Big|_{t_i=0} \qquad (2.7)$$

Formally we can write :

$$G^\beta(\psi \otimes f) = \sum_{n=0}^\infty \frac{j^n}{n!} \int_{H^*} d\mu^\varsigma(\Phi)(<\Phi, \psi \otimes f>)^n$$

$$= \sum_{n=0}^\infty \frac{j^n}{n!} S^\beta((\psi \otimes f),\ldots,(\psi \otimes f))$$

Thus, providing that the following estimate is valid :

$$|S^\beta((\tau_1,f_1),\ldots,(\tau_n,f_n))| \leq O((n!)^\gamma) \prod_{i=1}^n \|f_i\| \qquad (2.8)$$

for some $\gamma < 1$ and some continuous norm on $H_*$, then it is possible to reconstruct the thermal functional $\mathbf{G^\beta}$ from the corresponding Schwinger functions $S^\beta$.

Proposition 2.5

Let $\mathbf{S^\beta} = (S^\beta_n)$ be a system of multilinear, continuous functionals on the spaces $\otimes_{i=1}^n(C_r(K_\beta) \otimes H_*)$ and such that :

S(O) For some $\gamma < 1$ and some continuous norm on $H_*$

$$|S^\beta((\tau_1,f_1),\ldots,(\tau_n,f_n))| \leq O((n!)^\gamma) \prod_{i=1}^n \|\psi_i\| \} \|f_i\| \qquad (2.9)$$

S(1)  The system $\mathbf{S^\beta} = (S^\beta_n)$ is stochastically positive:
for any sequences $c_\alpha$ of complex numbers, $\alpha = 1:N$

$$\sum_{\alpha,\delta=1}^N c_\alpha^- c_\delta S^\beta_{\alpha+\delta}((\psi \otimes f)^\alpha, (\psi \otimes f)^\beta) \geq 0 \qquad (2.10)$$

where $(\psi \otimes f)^\alpha = ((\psi_1^\alpha, f_1^\alpha),\ldots(\psi_\alpha^\alpha, f_\alpha^\alpha))$.

S(2)   The system $S^{\beta} = (S^{\beta}_n)$ is reflection positive on $K_{\beta}$ :

for any sequences $c_{\alpha}$ of complex numbers, $\alpha = 1:N$

for any families $(\psi \otimes f)^{\alpha} = ((\psi_1^{\alpha}, f_1^{\alpha}), \ldots (\psi_{\alpha}^{\alpha}, f_{\alpha}^{\alpha}))$. with all $\psi$ supported on $[0, \beta/2]$

$$\sum_{\alpha,\delta}^{N} c_{\alpha}^{-} c_{\delta} \, S_{\alpha+\delta}^{\beta} \left( (\psi \otimes f)^{\alpha}, R(\psi \otimes f)^{\beta} \right) \geq 0 \qquad (2.11)$$

where $R(\psi \otimes f)^{\beta} = \left( \left( R\psi_1^{\beta}, f_1^{\beta} \right), \ldots \left( R\psi_{\beta}^{\beta}, f_{\beta}^{\beta} \right) \right)$
and R stands for the reflection operator.

S (3) the system $S^{\beta} = (S^{\beta}_n)$ is $K_{\beta}$ –shift invariant, β-periodic and R invariant.

Then there exists an unique thermal Green functional $\mathbf{G}^{\beta}$ such that the Schwinger functions of which is equal to $S^{\beta} = (S^{\beta}_n)$ .

Proof :

The only uniqueness of $\mathbf{G}^{\beta}$ might be unclear . However the estimates given by the S(0) are exactly of the form that guarantees that the corresponding moment problem for the corresponding path space of the underlying thermal process is unique .

The rest is more or less standard .   q.e.d.

β

## 3. From quasi-free structures to non- quasifree models.

A given β-KMS thermal Green functional $\mathbf{G}^{\beta}$ is called a quasi-free functional iff

$$\mathbf{G}^{\beta} (\psi \otimes f) = \exp(i(M(\psi \otimes f))) \ast \exp{-1/2} \, B(\psi \otimes f, \psi \otimes f) \qquad (3.1)$$

where M is a linear , continous functional on $C(K_{\beta}) \otimes H$ and B is a bilinear, continous (for simplicity) form on $C(K_{\beta}) \otimes H$ . In order to guarantee the properties **G(1)-G(3)** to be fulfilled we need the following properties for M and B to be obeyed:

**(qf0)**

there exists a selfadjoint , nonegative operator h acting in H such that for f,g $\in$ H

$$|B(f, g)| \leq \|f\| \ast \|g\| \qquad (3.2)$$

for some negative index Sobolev norm $\| \; \|$ connected with the operator h.

**( qf 1)**

the 2-form B is S-positive on $C(K_{\beta}) \otimes H$ i.e.

, i.e. for any sequences $c_{\alpha} \in \mathbb{C}$, $\psi_{\alpha} \in C(K_{\beta})$,

*Bose Matter models*

$f_\alpha \in H_*$, $\alpha=1{:}n$,

$$\sum_{\alpha,\gamma} c_\alpha^- * c_\gamma * B(\psi_\alpha \otimes f_\alpha, \psi_\gamma \otimes f_\gamma) \geq 0 \quad (3.3)$$

**(qf2)**
the 2-form B is R-positive on $C(K_\beta) \otimes H$ i.e.
, i.e. i.e. for any sequences $c_\alpha \in C$, $\psi_\alpha \in C([0,\beta/2])$, $f_\alpha \in H_*$, $\alpha=1{:}n$,

$$\sum_{\alpha,\gamma} c_\alpha^- * c_\gamma * G(\psi_\alpha \otimes f_\alpha, R\psi_\gamma \otimes f_\gamma) \geq 0 \quad (3.4)$$

where R is the reflection operator on $K_\beta$.

**qf (3)**
M and B are $K_\beta$ – shift invariant and R invariant.

In the case of of quasi free Green functional all the multitime Green are computed with the help of B and M.
For example :

$$G\big(((\tau_1,f_1))\big) = M(\delta_{\tau_1} \otimes f) \quad (3.5)$$

$$G\big(((\tau_1,f_1),(\tau_2,f_2))\big) =$$
$$M(\delta_{\tau_1} \otimes f_1)* M(\delta_{\tau_2} \otimes f_2) + B(\delta_{\tau_1} \otimes f_1, |\delta_{\tau_2} \otimes f_2) \quad (3.6)$$

Let us define $G^T$ as the corresponding truncated functions generating functional defined by the formula :

$$G^T\big(((\tau_1,f_1)\ldots(\tau_n,f_n))\big) = \sum_{\pi \in Par(1_n)} (|\pi|-1)!\,(-1)^{|\pi|-1} \prod_{\alpha=1}^{|\pi|} G((\tau,f)_{B_\alpha})) \quad (3.7)$$

Where: $1_n = (1,\ldots n)$, Par $(1_n)$ stands for the set of all partitions $\pi = (B_1,\ldots,B_k)$ of the set $1_n$ of indices and $((\tau,f)_{B_\alpha} = ((\tau_{1^\alpha}, f_{1^\alpha}),\ldots\ldots(\tau_{|B_\alpha|}, f_{|B_\alpha|^\alpha}))$.
The corresponding Schwinger functions are given by the following formulae :

$$S^\beta((\tau_1,f_1),\ldots,(\tau_n,f_n)) = \sum_{\pi \in 2Par(n)} \prod_{\alpha=1}^{k} S^{|B_\alpha|,T}(f_{B_\alpha}) \quad (3.8)$$

where $S^{1,T}$ corresponds to M and $S^{2,T}$ to the 2-form B respectively.
And the corresponding truncated Schwinger functions are easy to compute with result:

$S^{n,T} = O$ for n>2 and for n=1,2 as above, in ( 3.8)

A well known result can be easily deduced from the (3.7) and (3,* ).

Proposition 3.1
A β-KMS functional $\mathbf{G}^\beta$ is quasi-free iff all cumulants of the corresponding system $\mathbf{S}^\beta$ of order n>2 are equal to zero.
.

In the case of standard, nonrelativistic Bose matter the corresponding kernel B is denoted as $S^\beta_{0,\mu}$ and is given by the formula:

$$S^\beta_{0,\mu}(\psi \otimes f, \psi \otimes f) =$$
$$\int_{-\frac{\beta}{2}}^{\frac{\beta}{2}} ds \int_{-\frac{\beta}{2}}^{\frac{\beta}{2}} ds' \psi(s)\psi(s') < f | \frac{e^{-\beta|s-s'|h_\mu} + e^{-(\beta-|s-s'|)h_\mu}}{1-e^{-\beta h_\mu}} f > \quad (3.8)$$

where $h_\mu = -\Delta + \mu$, $-\Delta$ is the infinite volume selfadjoint, nonegative Laplace operator and $\mu > 0$ is the so called chemical potential.

Remark 3.1

Also the case of relativistic Euclidean scalar Bose field in the finite temperature β could be described by a similar formula (3.5), but with $h_\mu = \sqrt{-\Delta + \mu^2}$ with μ playing the role of mass.

Remark 3.2

The case of the critical phase of the standard nonrelativistic Bose matter is described as an properly chosen convex superposition of the corresponding pure phases thermal processes indexed by the parameters (α, ρ) corresponing, respectively to the U(1) symmetry braking in such phases and the density of the condensate contained. See our paper [10] for a detailed description of the corresponding thermal processes.

It is not difficult to verify that the kernel $S^\beta_{0,\mu}$ obeys the properies **qf(0) − qf(3)**. Therefore we can construct a standard, free Bose Gas (in the noncritical regime) Green functional

$$G_{0,\mu}(\psi \otimes f) = \exp{-1/2}\, S^\beta_{0,\mu}(\psi \otimes f, \psi \otimes f) \quad (3.9)$$

with the corresponding (eulidean) Green functions

$$G^\beta_{0,\mu}(((\tau_1, f_1) \ldots (\tau_n, f_n)) = \quad (3.10)$$
$$\prod_{1 \leq i \leq j \leq n} \exp{-1/2}\, S^\beta_{0,\mu}(\delta_{\tau_i} \otimes f_i, \delta_{\tau_j} \otimes f_j)$$

The properties of the corresponding thermal procesess were studied in details in our previous publications [13,19].

Let dP be a probability, Borel measure on the nonnegative reals $R^1_+$. Then, we can construct a new thermal kernel by the formula:

$$S^\beta_P = \int_{R^1_+} dP(\mu)\, S^\beta_{0,\mu} \quad (3.11)$$

**Proposition 3.1**
*Let dP be a Borel, probilistic measure on nonegative reals $R_+^1$. Then the thermal kernel ( 3.11 ) obeys the properties **qf0-qf3**.*
*Remark.*
In fact, the assumption that the measure dP is of probabilistic nature is not necessary in Proposition 3.1

From the above proposition it follows that defining the following thermal Green functional

$$\mathbf{G_P^\beta}(\psi \otimes f) = \exp\text{-}1/2\, S_P^\beta(\psi \otimes f, \psi \otimes f) \tag{3.12}$$

we have defined a functional which fulfills all the axioms **G1-G3** together with Schwarz type of regularity. This functional will be called a generalized Free Bose Matter functional. Detailed properties of the corresponding models will be presented elsewhere [10].

More interesting situations, from physical point of view are described by the following functional. Let dP be as before a probabilistic, Borel measure on $R_+^1$. Then we define the following thermal functional:

$$\mathbf{G^\beta}_\mathbf{P}(\psi \otimes f) = \int_{R_+^1} dP(\mu)\, \exp\text{-}1/2\, S_{0,\mu}^\beta(\psi \otimes f, \psi \otimes f). \tag{3.13}$$

The main result of this note is the following theorem.
**Theorem 3.1**
*For any probabilistic, Borel measure dP supported on $[e, \infty)$, for some $e > 0$ the functional $\mathbf{G^\beta}_P$ given by (3.13) obeys the axioms **G(1)-G(3)** together with regularity condition of Schwartz type.*
*The corresponding multitime (eulidean) Green functions of the system described by (3.13) are given by the formulae:*

$$G_P\big((\tau_1, f_1), \ldots, (\tau_n, f_n)\big) = \int_{R_+^1} dP(\mu) \prod_{1 \leq i \leq j \leq n} \exp - 1/2\, S_{0,\mu}^\beta(\delta_{\tau_i} \otimes f_i, \delta_{\tau_j} \otimes f_j). \tag{3.14}$$

**Proposition 3.3**
*Let the measure dP from Theorem 3.2 be supported at least on two atoms (points). Then the functional given by (3.14) defines β-KMS Green functional leading to some not quasi-free thermal KMS- structure.*
Proof :
Let us compute the truncated four times Green function:

$$G_P^T\big((\tau_1, f_1), \ldots, (\tau_4, f_4)\big) = \big(\int_{R_+^1} dP(\mu) \prod_{1 \leq i \leq j \leq 4} \exp - 1/2\, S_{0,\mu}^\beta(\delta_{\tau_i} \otimes f_i, \delta_{\tau_j} \otimes f_j)\big)^T$$

$$= G_P\big((\tau_1, f_1), \ldots, (\tau_4, f_4)\big) +$$
$$\sum_{\pi \in Par(1_4), \pi=(B_1,\ldots B_k)} (|\pi - 1|)!\, (-1)^{k-1} \prod_{\alpha=1}^k \big(\int dP(\mu) G_{0,\mu}^2((\tau, f)_{B_k})\big)$$

Where : $1_4 = (1,2,3,4)$, 2Par($1_4$) stands for the set of paritions of the set of indices $1_4$, where number of blocks is bigger then 1.

Also the truncated Schwinger functions called cumulants of the functional $\mathbf{G}^\beta$ are all nonzero for n >2 , providing the measure dP is supported on the set consisting of at least two points. This can be observed from the following formulae:

For simplicity only of our exposition let us take

dP ( μ )= ½ ( δ ($\mu - \mu_1$) +δ ($\mu - \mu_2$)) , $\mu_1 \neq \mu_2$

Then the 2-point Schwinger function is given as

$S_P^2((\tau_1, f_1), (\tau_2, f_2)) = \frac{1}{2} S_{0,\mu_1}^\beta ((\tau_1, f_1), (\tau_2, f_2)) + \frac{1}{2} S_{0,\mu_2}^\beta ((\tau_1, f_1), (\tau_2, f_2))$

(3.8)

Let us compute the 4- point truncated Schwinger function:

$$S_P^{4,T}((\tau_1, f_1), \dots, (\tau_4, f_4))$$
$$= S_P^4(((\tau_1, f_1), \dots, (\tau_4, f_4))$$
$$- \sum_{\pi=(B_1, B_2) \in 2Par(4)} S_P^2((\tau, f)_{B_1}) S_P^2((\tau, f)_{B_2}) =$$

$= \frac{1}{4} S_{0,\mu_1}^4((\tau_1, f_1), \dots, (\tau_4, f_4)) + \frac{1}{4} S_{0,\mu_2}^4((\tau_1, f_1), \dots, (\tau_4, f_4)) -$
$\frac{1}{4} \sum_{\pi=(B_1, B_2) \in 2Par(4)} S_{0,\mu_1}^2((\tau, f)_{B_1}) S_{0,\mu_2}^2((\tau, f)_{B_2}) + S_{0,\mu_1}^2((\tau, f)_2) S_{0,\mu_2}^2((\tau, f)_{B_1}))$

*(3.9)*

which is definitely nonzero for , but equal to zero in the case $m_1 = m_2$.
Similarly, one can see in an explicite form that the higher order cumulants of the Schwinger functional are all nonzero for $m_1 \neq m_2$.

Thus providing that the measure dP is not supported in one point we are seeing that four point cumulant of $\mathbf{S}_P$ is always nonzero. Then also the higher order cumulants are non zero .

q.e.d.

The necessary and sufficient condition for being quasi-free is that all higher order ,cumulants of $\mathbf{S}^\beta$ for n>2 , cumulants are equal to zero. Then because the corresponding truncated multitime functions does not vanish for n>2 some non quasi-free KMS-structures on the Weyl algebra W ( $L_2$ ( $R^d$ )) can be constructed from the system of multi-time Green functions as described in [18]. The study of the corresponding interacting models of Bose matter will be presented elsewhere [10].

*Bose Matter models*